\documentstyle[12pt]{article}
\begin{document}
\tolerance 6000
\hbadness 60007
\title{A SIMPLE ACTION FOR A FREE ANYON }
\author{D. Dalmazi\thanks{e-mail: dalmazi@feg.unesp.br}~~ 
and A. de Souza Dutra\thanks{e-mail: dutra@feg.unesp.br} \ 
\\UNESP - Campus de Guaratinguet\'a - 
DFQ \\Av. Dr. Ariberto Pereira da Cunha, 333 - CEP 12500-000 \\
Guaratinguet\'a - SP - Brasil}
\date{}
\maketitle
\thispagestyle{empty}
\begin{abstract}
By studying classical realizations of the sl(2,{$\Re$}) algebra in a two
dimensional phase space $(q,\pi )$, we have derived a continuous family of
new actions for free anyons in 2+1 dimensions. For the case of light-like
spin vector $\left( S_\mu S^\mu =0\right) $, the action is remarkably
simple. We show the appearence of the Zitterbewegung in the solutions of the
equations of motion, and relate the actions to others in the literature at
classical level.
\end{abstract}

\newpage

\section{Introduction}

One of the main motivations in the physics of particles with fractional
statistics in $2+1$ dimensions (anyons [1]) is to address the questions of
fractional quantum hall effect and high temperature superconductors [2,3]
which are presumably planar effects. Typically one makes use of a
statistical field $A_{\mu}$ of the Chern-Simons type, coupled to some matter
field whose statistics will be changed by $A_{\mu}$. One interesting
question about this coupling is whether some interaction is left for the
anyon besides the change of statistics. This question suggests the search
for actions for free anyons as elementary relativistic particles. In
particular, from the point of view of point particles, there are already
several works in the literature sugesting different actions for anyons
[4-12]. Care should be taken with the fact that in $2+1$ dimensions the
irreps of the Lorentz group are finite only for integer or half-integer
spins. Thus, for fractional spins we have to deal with infinite dimensional
representations.

There are basically two kinds of point-particle actions for anyons in the
literature. Since anyons have the same number of degrees of freedom as
massive spinless particles, they can be described by the same space-time
variables $\,(x_{\mu},p_{\nu})\,$, see \cite{schon,jn} and also \cite{chou}.
 The price paid in this minimal
formulation is the appearance of non canonical brackets among the
space-time variables ($\{x_{\mu},x_{\nu}\}\ne 0 $)
and anomalous spin algebra ($\{ S_{\mu},S_{\nu}\}=0 \, $ instead of $\{
S_{\mu},S_{\nu}\}=\, \epsilon_{\mu\nu\alpha}S^{\alpha}$) besides 
the lack of
explicit Lorentz covariance. On the other hand, by adding extra variables,
which must be subtracted afterwards, one can surmount these difficulties and
even write explicitly Lorentz covariant actions , see e.g. \cite{ijpl} ,
\cite{gosh}.
\footnote{We remark that in \cite{gosh} there is one
exceeding degree of freedom}. However, in these cases one has to face rather
complicated Lagrangians. In the literature there are also
extended formulations which are not explicitly covariant , e.g. ,
\cite{bala},\cite{minipl}.

In this work we derive the minimal extension of the space-time phase space
such that the canonical structure of the space-time and the canonical spin
algebra are preserved. In order to circumvent the problem of lack of
translation invariance raised up in \cite{plb} we use $\dot
p_{\mu}=0$ replacing $\ddot x_{\mu}=0$ as a definition of a free particle
which permits the appearance of the Zitterbewegung that will indeed
show up. We end up with a family of rather interesting actions for free
anyons which is specially simple when the 
spin is light-like ($S^{\mu}S_{\mu}=0$).

In section 2 we briefly review the covariant formulation of \cite{ijpl}
showing the existence of a gauge 
\cite{ijpl} showing the existence of a gauge
condition where the canonical structure of the space-time and the spin
variables are preserved. In section 3 we study classical realizations of the
sl(2,$\Re $) spin algebra in a two dimensional phase-space deducing the
family of Lagrangians cited before and showing the presence of the
Zitterbewegung in the classical solutions. Our 
Lagrangian , for the case $S^{\mu}S_{\mu}=0$ is directly
related to the covariant approach of \cite{ijpl} through a convenient gauge
fixing followed by a Darboux transformation as shown in section 4. Section 5
contains some final comments and perpectives.

\section{Covariant Description of Anyons}

The Poincar\`e algebra in 2+1 dimensions is given by

$$
[J_\mu ,J_\nu ]=i\epsilon _{\mu \nu \alpha }J^\alpha 
$$

\begin{equation}
\label{II-1}[J_\mu ,P_\nu ]=i\epsilon _{\mu \nu \alpha }P^\alpha 
\end{equation}

$$
[P_\mu ,P_\nu ]=0 
$$

\noindent where $J_\mu $ are the dual components of the total angular
momentum $J_\mu =\,\epsilon _{\mu \nu \alpha }J^{\nu \alpha }$ and we use $\eta
_{\mu \nu }=\left( +,-,-\right) $, $\epsilon _{012}\,=+1$. Classically the
Poincar\`e algebra can be realized in terms of Poisson brackets with the
identifications:

\begin{equation}
\label{II-2}J_\mu =\,\epsilon _{\mu \nu \alpha }x^\nu p^\alpha +S_\mu (q_i,\pi
_j)\,\,\,,\,\,\,P_\mu =p_\mu \,, 
\end{equation}

\noindent where we have introduced an extended phase space which besides the
usual canonical variables $\left( x_\mu ,p_\nu \right) $, necessary to
describe a spinless particle, posses also extra canonical variables $\left(
q_i,\pi _j\right) $; $i=1,...,N$, used here to describe the dual spin
components $S_\mu $. The only non vanishing brackets are $\left\{ q_i,\pi
_j\right\} =\delta _i^j$; $\left\{ x_\mu ,p_\nu \right\} =\delta _\mu ^\nu $%
. Moreover $S_\mu $ satisfy the canonical spin algebra (sl(2,$\Re $)):

\begin{equation}
\label{II-3}\left\{ S_\mu ,S_\nu \right\} =\,\epsilon _{\mu \nu \alpha }
S^\alpha. 
\end{equation}

The Poincar\`e algebra (\ref{II-1}) contains two quadratic Casimir
invariants $P^2=p_\mu p^\mu $ and $J\cdot P=S_\mu p^\mu $ which specify the
mass and the helicity of the anyon respectively. In order to assure that the
physical states of the anyon belong to an irreducible representation of
Poincar\`e algebra with given mass ($m$) and helicity ($\alpha $) we
associate the Casimir invariants with two first class constraints following 
\cite{jn}:

$$
\phi _1=p^2-m^2\approx 0\,\,\,,\,\,\,\phi _2=S\cdot p+\alpha m\approx 0 
$$

\noindent where $\alpha $ is a given real constant.

Another important ingredient when constructing a relativistic model for an
anyon, is the correct counting of degrees of freedom. As shown in 
\cite{jn}\cite{ohnu} the anyon posses, irrespective of $\alpha $, just one
polarization state like a spinless particle. Therefore the extra variables $%
\left( q_i,\pi _j\right) $ must be subtracted by means of suitable
additional constraints. Since $\phi _2$ is first class, the extra
constraints $\Phi $$_A,\,A=1,2,...,l$ must be such that we get rid of all
the remaining $2N-2$ variables among $\left( q_i,\pi _j\right) $. The number
`$l$ ' of such constraints depend upon their class, i. e., first or second
class.

Considering all those ingredients and bearing in mind that first quantized
theories of relativistic particles have vanishing canonical Hamiltonian, the
general form of an anyon total Hamiltonian is given by \cite{plb}:

\begin{equation}
\label{II-4}H=\frac e2\,\phi _1\,+\,\sigma \,\phi _2\,+\,\lambda ^A\Phi _A 
\end{equation}

\noindent where $e$, $\sigma $ and $\lambda ^A$ are non-physical arbitrary
functions of time. There are many choices for the extra variables $\left(
q_i,\pi _j\right) $, and also different possibilities for the constraints $%
\Phi $$_A$ which will lead in general to different parameterizations of the
free anyon action, which should be physically equivalent. Probably the
minimal way to describe the vector $S_\mu $ in an explicitly covariant form
is to introduce a vector $n_\mu $ and its conjugate momentum $\pi _\mu $,
with the only non vanishing brackets $\left\{ n_\mu ,\pi _\nu \right\} =\eta
_{\mu \nu }$.

Now we can write $S_\mu =\,\epsilon _{\mu \nu \alpha }n^\nu \pi ^\alpha $, 
which
satisfy the sl(2,$\Re $ ) algebra (\ref{II-3}). In the rest of this section
we will concentrate on this minimal covariant extension. Next step is to
choose the constraints $\Phi $$^A$ in order to eliminate four of the extra
variables $\left( n_\mu ,\pi _\nu \right) $ since two of them will be
already eliminated by $\phi _2$. It is clear that, the less constraints we
have in (\ref{II-4}) the simpler will be the Lagrangian obtained by the
inverse Legendre transformation of $H$. Thus, the simplest choice for $\Phi $%
$^A$ corresponds to two first class constraints $\Phi $$_1,\,\Phi _2$. The
requirement of Lorentz covariance and the assumption of linear independence
of the constraints demands that $\Phi $$_i$ be scalars. The scalars should
not involve $x^\mu $ since $\dot p_\mu =0$ for a free anyon. Therefore $\Phi 
$$_i$ must be functions of $\pi ^2,n^2,\pi \cdot n,p\cdot n,$ and $\pi \cdot
p$. We have verified that the choice $\Phi $$_1=\pi \cdot n\,\approx \,0$, $%
\Phi $$_2=p\cdot n\,\approx \,0$ leads to the simplest Lagrangian, which is
given by 
\begin{equation}
\label{II-5}L=-m\frac{\epsilon _{\alpha \beta \gamma }\dot x^\alpha n^\beta 
\dot
n^\gamma }{\left[ \dot n^2n^2-\left( n\cdot \dot n\right) ^2\right] ^{\frac
12}}+\frac \alpha {n^2}\left[ \dot n^2n^2-\left( n\cdot \dot n\right)
^2\right] ^{\frac 12}. 
\end{equation}

\noindent From this Lagrangian we get four primary first class
constraints $\phi _1,\phi _2,\Phi _1,\Phi _2$ and no further ones appear,
such that the Hamiltonian is of the form (\ref{II-4}):

\begin{equation}
\label{II-6}H=\frac e2\,\left( p^2-m^2\right) +\sigma \,\left( S\cdot
p+\alpha m\right) +\lambda _1\pi \cdot n+\lambda _2\,\,p\cdot n. 
\end{equation}

\noindent with $S_\mu =\,\epsilon _{\mu \nu \alpha }n^\nu \pi ^\alpha $.
According to our knowledge this theory was first suggested in \cite{ijpl}
where it was quantized in a gauge independent way. However to get closer to
the physical degrees of freedom, it is important to further fix the gauge of
the non universal constraints $\Phi $$_1,\Phi _2$ while keeping the
essential constraints $\phi _1,\phi _2$ first class, in analogy to the
pseudo-classical description of relativistic spinning particles of \cite{ber}%
\cite{div}, where the local 1D-supersymmetry constraint plays the r\^ole of
the $\phi _2$ constraint giving rise to the Dirac propagator of spin $\frac
12$ particles.

After introducing the gauge conditions $\chi _i=0,\,i=1,2$ the constraints $%
\Phi $$_i$ will turn into second class. In order that $p^\mu $ be still
interpreted as translation generators, i. e., $\left\{ x_\mu ,p_\nu \right\}
_{DB}=\eta _{\mu \nu }$ the gauge conditions must be $x^\mu $ independent.
Moreover, Poincar\`e covariance restricts $\chi _i$ to be functions of the
scalars $\pi ^2,n^2$ and $\pi \cdot p$. There are, among several
possibilities, two specially interesting gauges, namely

$$
\chi _1^I=\pi \cdot p=0\,,\,\chi _2^I=n^2-a=0\,, 
$$

\begin{equation}
\label{II-7}\chi ^{II}_1=\pi ^2-b=0\,,\,\chi ^{II}_2=n^2-a=0\,, 
\end{equation}

\noindent with $a,b\,\in \,\Re $ and $a<0$. In the case (I) $S_\mu $
and $p_\mu $ become parallel which makes the spin algebra anomalous: $%
\left\{ S_\mu ,S_\nu \right\} _{DB}^I=0$ and originates non-commuting
coordinates ($\left\{ x_\mu ,x_\nu \right\} _{DB}^I\neq 0$), mixing spin and
space-time $\left\{ x_\mu ,S_\nu \right\} _{DB}^I\neq 0$. For the case (II)
none of those problems appear. In particular, the canonical structure of $%
\left( x_\mu ,p_\nu \right) $ small phase space is maintained and the sl(2,{$%
\Re $ }) spin algebra is not broken $\left( \left\{ S_\mu ,S_\nu \right\}
_{DB}^{II}=\,\epsilon _{\mu \nu \alpha }S^\alpha \right) $ 
which certainly simplifies the
quantization. In both gauges the Poincar\`e 
covariance is not spoiled.

The above example teaches us that the spin algebra in a covariant theory for a
free anyon is a gauge dependent quantity, and it is always possible to
choose a gauge where the canonical sl(2,{$\Re $ }) spin algebra (\ref{II-3})
is obeyed which is, as we have seen, a natural gauge choice avoiding
complicate Dirac brackets. The sl(2,{$\Re $ }) spin algebra will guide us in
the next section in finding a new rather simple action for a free anyon.

\section{A Simple Action for an Anyon}

According to the last section analysis of the degrees of freedom, if we
extend the spinless particle phase space $\left( x_\mu ,p_\nu \right) $
introducing only one couple of conjugated variables $\left\{ q,\pi \right\}
=1$, no extra constraints $\Phi ^A$ will be necessary besides the first
class constraints $\phi _1=p^2-m^2$ and $\phi _2=S\cdot p+\alpha m\,$ such
that the Hamiltonian (\ref{II-4}) becomes:

\begin{equation}
\label{III-1}H\left( p^\mu ,q,\pi \right) =\frac e2\left( p^2-m^2\right)
\,+\,\sigma \left( S_\mu \left( q,\pi \right) p^\mu +\alpha m\right) . 
\end{equation}

In this minimally extended phase space we suppose that $S_\mu (q,\pi )$ are
analytical functions of $q,\,\pi $ and satisfy the canonical sl(2,{$\Re $ })
spin algebra (\ref{II-3}). The simplest Ansatz for $S_\mu (q,\pi )$
compatible with those hypotheses is a linear function of $\pi $:

\begin{equation}
\label{III-2}S_\mu =f_\mu (q)\pi \,+\,g_\mu (q), 
\end{equation}

\noindent with $f_\mu (q)$ and $g_\mu (q)$ analytic functions of $q$ to be
fixed by the algebra (\ref{II-3}), that leads to the nonlinear differential
equations: 
\begin{equation}
\label{III-3}f_\mu =\,\epsilon _{\mu \alpha \beta }f^{\prime \alpha }f^\beta 
\,, 
\end{equation}

\begin{equation}
\label{III-4}\,g_\mu =\,\epsilon _{\mu \alpha \beta }g^{\prime \alpha }f^\beta
\,\,, 
\end{equation}

\noindent where $f^{\prime }(q)=df/dq$. After some algebra we show that the
general solution to (\ref{III-4}), compatible with (\ref{III-3}) is

\begin{equation}
\label{III-5}g_\mu (q)\,=r(q)\,f_\mu \,+\,s\,f\,_\mu ^{\prime } 
\end{equation}

\noindent where $r(q)$ is an arbitrary function and $s$ is an arbitrary real
constant. Back in (\ref{III-2}) we get $S_\mu =\,(\pi +r(q))f_\mu
\,+s\,f_\mu ^{\prime }$. Since $(q,\pi )\longrightarrow (q,\pi +r(q))$ is a
canonical transformation that changes the Lagrangian by a total time
derivative, we can set $r(q)=0$ without loss of generality from the physical
point of view, such that in general

\begin{equation}
\label{III-6}S_\mu =\pi \,f_\mu \,+\,s\,\,f_\mu ^{\prime }. 
\end{equation}
\noindent From (\ref{III-3}) $f_\mu $ must satisfy in particular:

\begin{equation}
\label{III-7}f^2=\,0\,\,;\,\,f^{\prime 2}=\,-1\,\,;
\,\,f^{\prime \prime }\cdot f\,=\,1. 
\end{equation}

\noindent Then we have

\begin{equation}
\label{III-8}S^2=-s^2. 
\end{equation}

\noindent Notice that the appearance of another arbitrary constant ($s$)
besides $\alpha $ and $m$ is not surprisingly since, due to the sl(2,{$\Re $ 
}) algebra (3),
it is easy to deduce
that the Casimir of the spin algebra $S^2$ is an observable in the
Poincar\`e multiplet, i. e., $\left\{ S^2,J_\mu \right\} =0=\left\{
S^2,P_\mu \right\} $. In the last section covariant example
 the same arbitrary
constant appear through the gauge conditions ($\chi _i^{II}=0$),
resulting in : $S^2=a\,b$.

Coming back to the differential equations (\ref{III-3}), although we do not
know their general solution, the simplest analytical solution 
(see identities (\ref{III-7})) is a second
order polynomial
\footnote{ When quantized ($\pi \sim d/dq$) these realizations correspond to
the differential operators commonly used for the so(2,1) spectrum generating
algebra, see for example \cite{bar}. } :

\begin{equation}
\label{III-9}f_\mu ^{(2)}=a_\mu \,\,q^2+b_\mu \,\,q+c_\mu
\,, 
\end{equation}

\noindent where for instance, $a_\mu =(1,0,-1);\,\,b_\mu =(-1,1,1);\,\,c_\mu
=\frac 12(1,-1,0)$. It is worthwhile to point out that we do not really need
to know the general solution for $f_\mu (q)$ since any other realization $%
\tilde S_\mu (\tilde q,\tilde \pi )$, with the same Casimir ($\tilde S^2=S^2$%
) will be physically equivalent to (\ref{III-9}). Indeed,
imposing $\tilde S_\mu (\tilde q,\tilde \pi )=S_\mu ^{(2)}(q,\pi
)=\pi \,\,f_\mu ^{(2)}(q)\,+\,s\,\,f_\mu ^{\prime (2)}$, two of these
equations, for instance $\tilde S_1=S_1^{(2)}$ and $\tilde S_2=S_2^{(2)}$
will define two different relationships
of the kind $q^{(a)}=q^{(a)}(\tilde
q,\tilde \pi )\,\,;\,\pi ^{(a)}=\pi ^{(a)}(\tilde q,\tilde \pi )\,\,(a=1,2)$.
The identification $\tilde S_\mu =S_\mu ^{(2)}$ implies of course
the identification of the corresponding Casimirs $\tilde
S^2=(S^{(2)})^2 $ which guarantees that $\tilde S_0=\pm \,S_0$. The two
signs correspond to the two different possibilities $q^{(a)},\pi ^{(a)}$ .
Therefore there will always be a canonical transformation bringing $\tilde
S_\mu (\tilde q,\tilde \pi )$ to the form $S_\mu ^{(2)}(q,\pi )$. Since this
proof assumes no hypothesis whatsoever on the form of $\tilde S_\mu (\tilde
q,\tilde \pi )$ it is clear now that any other more involved realization
of sl(2,{$\Re $ }) algebra than (\ref{III-2})
with $\tilde S^2 =-s^2  $  it is an unnecessary
complication. Thus, we conclude that the Hamiltonian (\ref{III-1}) 
is unique up to canonical transformations and it leads,
by an inverse Legendre transformation, to a unique Lagrangian up to a total
time derivative : 
\begin{equation}
\label{III-10}L=\,p^\mu \dot x_\mu \,+\,\pi \,\dot q\,-\,\frac e2\left(
p^2-m^2\right) \,-\sigma \,\left( \pi \,f\cdot p\,\,+\,s\,\,f^{\prime }\cdot
p\,+\,\alpha m\right) . 
\end{equation}

The Legendre transformation is singular due to the Lagrangian constraint $%
\sigma \,\,f\cdot \dot x\,=\,e\dot q\,$ but after a careful analysis, and
eliminating some auxiliary fields we finally get: 
\begin{equation}
\label{III-11}L_s=\,\frac 1{2e}\left( \dot x_\mu -\frac{s\,\dot q\,f_\mu
^{\prime }}{f\cdot \dot x}\right) ^2\,+\,\frac{e\,m^2}2\,-\,\frac{\alpha
\,m\,e\,\dot q}{f\cdot \dot x}. 
\end{equation}

\noindent The quantity $f_\mu (q)$ must be a solution of the equation (\ref
{III-3}), the Lagrangian $L_s$ can be written in a polynomial form
reintroducing the auxiliary fields. From $L_s$ one derives
two primary constraints $\pi _e\approx 0;\,\pi \,f\cdot
p\,\,+\,s\,\,f^{\prime }\cdot p\,+\,\alpha m=\phi _2\approx 0$, and the
secondary one $\phi _1=p^2-m^2\approx 0$; as expected. The quantities
$S^2,\,S\cdot p$ and $p^2$ are dynamically independent from which it 
follows that $L_s$ represents a continuous family of 
Lagrangians describing
anyons with helicity $\alpha $, mass $m$ and $S_\mu S^\mu =-s^2$. Specially
simple is the case where $S_\mu $ is light-like ($s=0$) 
\begin{equation}
\label{III-12}L_0=\,\frac{\dot x^2}{2e}\,+\,\frac{e\,m^2}2\,-\,\frac{\alpha
\,m\,e\,\dot q}{f\cdot \dot x}. 
\end{equation}

\noindent It is remarkable that we just have to add the last term above with 
$f_\mu (q)$ a solution of (\ref{III-3}) like e. g. (\ref{III-9}), in order
to obtain an anyon from a relativistic spinless massive particle. The
Lagrangians (\ref{III-11}) and (\ref{III-12}) are the main results of this
work. The Lagrangian $L_0$ is the simplest that we know describing a free
anyon with the correct counting of degrees of freedom and that preserves the
sl(2,{$\Re $ }) spin algebra (\ref{II-3}) and the canonical structure of the
space-time variables ( $\left\{ x_\mu ,x_\nu \right\} =0=$ $\left\{ p_\mu
,p_\nu \right\} ;\,$ $\left\{ x_\mu ,p_\nu \right\} =\eta _{\mu \nu }$).

Due to the fact that $f_\mu $ is a function of one variable $f_\mu =f_\mu
(q) $ which is a solution of equation (\ref{III-3}), 
the Lagrangians (\ref{III-11}) and (\ref{III-12}), despite their
appearance, are not Lorentz covariant. However, by performing a Lorentz
rotation with constant parameter $k^\mu \,(k^\mu =\,\epsilon ^{\mu \nu \alpha
}k _{\nu \alpha })$: 
\begin{equation}
\label{III-13}\delta x_\mu =\,\epsilon _{\mu \alpha \beta }k^\alpha x^\beta
\,\, ; \,\,\delta q=k^\mu f_\mu (q). 
\end{equation}

\noindent it is easy to derive (using equation (\ref{III-3})): 
\begin{equation}
\label{III-14}\delta L_s=\frac d{dt}\left( -s\,f^{\prime }\cdot k\right) . 
\end{equation}

\noindent Therefore the action $S=\int dt\,L_s$ is Lorentz, actually
Poincar\`e, invariant. The Noether theorem will lead to the conserved
quantities $J_\mu $ and $P_\mu $ of (\ref{II-2}) with $S_\mu $ given in (\ref
{III-6}). Those charges close the Poincar\`e algebra in terms of Poisson
brackets. Thus, at least from the classical point of view, the Lagrangian $%
L_s$ describes perfectly well defined relativistic invariant theories of 
free anyons, in a similar way to what happens to the Floreanini-Jackiw 
\cite{fj} action for chiral bosons or the Schwarz-Sen \cite{ss} action for
electrodynamics. The loss of explicit Lorentz covariance was the price that
we have paid for working in a small configuration space $L_s(\dot x_\mu
,q,\dot q)$ instead of $L(\dot x_\mu ,n_\mu ,\dot n_\alpha )$ of the
covariant theory of previous section. In the next section we will return to
the relation between our approach and the explicitly covariant one.

We finish this section analyzing the equations of motion of $L_s$ which are
equivalent to the Hamilton equations of 
\begin{equation}
\label{III-15}H_s=\frac e2\left( p^2-m^2\right) \,+\,\sigma \,\left( S_\mu
(q,\pi )p^\mu +\alpha m\right) \,+\,\mu \,\pi _e, 
\end{equation}

\noindent where $S_\mu (q,\pi )$ is given in (\ref{III-6}). It is easy to
derive that 
\begin{equation}
\label{III-16}\pi =-\frac{(sf^{\prime }\cdot p+\alpha m)}{f\cdot p}%
\,\,\,;\,\,\,\dot q(t)\,=\,\sigma f^\mu (q)p_\mu 
\end{equation}
\begin{equation}
\label{III-17}\dot x_\mu =e\,p_\mu \,+\,\sigma \,S_\mu \,\,;\,\,\,\,\,\,\dot
p_\mu =0, 
\end{equation}

\noindent besides, the constraints $p^2-m^2=0$ and $\pi _e=0$. Even in the
simplest case when $f_\mu (q)=f_\mu ^{(2)}(q)$ the equation for $q(t)$ is a
bit complicated, but writing it in terms of $S_\mu $ we have: 
\begin{equation}
\label{III-18}\dot S_\alpha =\sigma \,\,\epsilon _{\alpha \mu \gamma }p^\mu
S^\gamma , 
\end{equation}

\noindent consequently: 
\begin{equation}
\label{III-19}\ddot S_\alpha -\frac{\dot \sigma }\sigma \,\dot S_\alpha
\,+\,m^2\sigma ^2S_\alpha \,=\,-\alpha \,m\sigma ^2p_\alpha . 
\end{equation}

\noindent In a gauge where $\dot \sigma =0=\dot e$ we have the solution: 
\begin{equation}
\label{III-20}S_\alpha (t)=-\frac \alpha mp_\alpha +\frac \sigma \omega
\left( p^2c_\alpha -p\cdot c\,p_\alpha \right) \cos \left( \omega t\right)
+\epsilon _{\alpha \beta \gamma }p^\beta c^\gamma \sin \left( \omega t\right) 
\end{equation}
$$
x_\mu (t)=\,x_\mu (0)+\left( e-\frac \alpha m\sigma \right) p_\mu t+\left(
\frac \sigma \omega \right) ^2\left( p^2c_\alpha -p\cdot c\,p_\alpha \right)
\sin \left( \omega t\right) + 
$$
\begin{equation}
\label{III-21}-\frac \sigma \omega \left( \epsilon _{\alpha \beta \gamma }p^\beta
c^\gamma \right) \cos \left( \omega t\right) , 
\end{equation}

\noindent where $\omega =m\mid \sigma \mid $ and $c_\alpha $ is a constant
vector with only  one independent component
which is related to $q(0)$. Therefore, in agreement with 
\cite{plb}, in order to interpret $x_\mu (t)$ as center of mass
coordinates of a free rigid body (\"x$_\mu =0$) $S_\alpha $ and $p_\alpha $
must be parallel but in general this is not the case and
the oscillating motion (Zitterbewegung) will take place along the orthogonal
directions to $p_\mu $.

\section{Connection to other Lagrangians}

The simplicity of $L_o$ when compared to other approaches like the covariant
one of \cite{ijpl} rises up the question about the link
between these theories. It will be seen that there is a rather interesting
and direct connection between these formulations. In order to compare the
covariant approach sketched in section 2 to $L_o$ we have to fix the gauge
of the constraints $\pi \cdot n=0$ and $p\cdot n=0$ (see (\ref{II-6})) such
that the $SL(2,\Re )$ spin algebra is preserved and furthermore $%
S^2=(\epsilon _{\mu \nu \alpha }n^\alpha \pi ^\alpha )^2=n^2\pi ^2-\left(
\pi \cdot n\right) ^2=0$. The gauge conditions $\chi _1=\pi ^2=0$\thinspace
; $\chi _2=n^2-a=0$ ($a$ is a negative constant) will enforce the two
previous requirements. We end up with four second class constraints: 
\begin{eqnarray}
n^2 - a =  0 &; &\pi^2=0  \\
p\cdot n =0 & ; &\pi \cdot n = 0 \nonumber\\
\end{eqnarray} 
Those can be explicitly solved for, e.g. , $n_\mu $ and $\pi _0$ as
functions of $\pi _1$ and $\pi _2$: 
\begin{equation}
n_\mu =\pm \sqrt{-a}\frac{\epsilon _{\mu \nu \alpha }p^\nu \pi ^\alpha }{\pi
\cdot p}\,\,\,;\,\,\pi _0=\pm \sqrt{\pi _1^2+\pi _2^2} 
\end{equation}
which imply $\,S_\alpha =\pm \sqrt{-a}\,\pi _\alpha \,$ . Henceforth $%
\,n_\mu \,$ and $\,\pi _{0\,}$ will be understood as (31). The 
$\,sl\left( 2,\Re \right) $ algebra for $\,S_\mu \,$ follows from the Dirac
brackets: 
\begin{equation}
\left\{ \pi _\alpha ,\pi _\beta \right\} _{DB}\,=\, - \, \frac{\epsilon
_{\alpha \beta \gamma }S ^\gamma }{a} \,=\,\pm \frac{\epsilon
_{\alpha \beta \gamma }\pi ^\gamma }{\sqrt{-a}} \,
\end{equation}
Inserting the solutions (31) into the first order form of the covariant
Lagrangian of section 2 , namely, 
\begin{equation}
L\,=\,p\cdot \dot x\,\,+\,\pi \cdot \dot n\,-\,\frac e2\left( p^2-m^2\right)
\,-\,\sigma \left( S\cdot p\,+\alpha m\right) \,-\,\lambda _1\pi \cdot
n\,-\,\lambda _2p\cdot n 
\end{equation}
we derive, after elimination of $\,p_\mu $\thinspace , 
\begin{equation}
\tilde L=\frac{\left( \dot x_\mu -\sqrt{-a}\sigma \,\pi _\mu \right) ^2}{2e}%
\,+\,e\,m^2\,-\,\alpha \,m\,\sigma \,+\,\frac{\sqrt{-a}}{\pi \cdot \dot x}%
\,\epsilon _{\alpha \beta \gamma }\dot x^\alpha \dot \pi ^\beta \pi ^\gamma 
\end{equation}

\noindent It can be shown that this intermediate Lagrangian corresponds
exactly to the one written in formula (4.2) of \cite{minipl} for the case of
lightlike spin \thinspace $\left( S^\mu S_\mu =0\right) $ where the
following assignments are understood:\thinspace $\sigma =v\,;\,\sqrt{-a}\pi
_{\mu \,}\,=\,j_\mu \,$. We emphasize that $\tilde L$ is not explicitly
Lorentz covariant despite its appearance since $\,\pi _0=\pm \sqrt{\pi
_1^2+\pi _2^2}\,$ and that makes it more complicated than it looks like. Now
a comment is in order; The hamiltonian reduction from $\,\left( n_\mu ,\pi
_\nu \right) \,$ to $\left( \pi _1,\pi _2\right) $ keeps the constraint $%
\,\phi _2=S\cdot p\,+\,\alpha m\approx 0\,\,$first class ,thus it should
furnish a similar Lagrangian to $L_0$. 
The fact that $\tilde L$ depends on two
extra dynamical variables besides \thinspace $x_\mu $ ($L=L\left( x_\mu ,\pi
_1,\pi _2\right) )\,,$ instead of just one like $\,L_0=L_0\left( x_\mu
,q\right) $ is a consequence of the Dirac brackets (32) which show that $\pi
_1$ and $\,\pi _2$ are not canonically conjugated variables like the couple $%
\,\left( q,\pi \right) \,$ of section 3. That is one of the sources of
complications in the intermediate Lagrangian $\tilde L$. Indeed, comparing
the definition of spin of our approach for $s=0$ with the case
of $\tilde L$ ($S_\mu =\sqrt{-a}\pi _\mu $) we have the
identification: 
\begin{equation}
\pi _\mu =\frac{\pi f_\mu \left( q\right) }{\sqrt{-a\,}} 
\end{equation}
\noindent which when introduced in $\tilde L$ reduces it to 
precisely $\,L_0\,$
. Therefore we conclude that $L_0\,$ can be obtained directly from the
covariant theory reviewed in section 2 through a gauge fixing
procedure followed by the non canonical transformation (Darboux
transformation) (35) that brings the non canonical couple $\,\left( \pi
_1,\pi _2\right) \,$ into the canonical one $\,\left( q,\pi \right) $.
Dynamically the situation is that on one hand, after the gauge fixing, both
 $\,\pi _1\,$and $\,\pi _{2\,}$ propagate while 
after the Darboux transformation
(35) on the other hand only $\,q\,$ propagates and $\pi \,$ can be
eliminated by its equation of motion. The extra constraints which certainly
appear in the first case are responsible for the match of the degrees of
freedom in both cases.

Last we comment upon the case $\left( S^\mu S_\mu =-s^2\neq 0\right) \,.$
Reintroducing auxiliary fields, $\,L_s\,$ can be rewritten as: 
\begin{equation}
L_s\,=\,\frac{\left( \dot x_\mu -\sigma S_\mu \right) ^2}{2e}\,+\,\frac{em^2}%
2\,-\,\alpha m\sigma \,+\,\pi \dot q 
\end{equation}
where $\,S_\mu =\pi f_{\mu \,}+sf_\mu ^{\prime }\,.$ Comparing with the
intermediate Lagrangian of \cite{minipl} we are led to another
Darboux transformation similar to the previous one,
\begin{equation}
j_\mu =\pi f_\mu \left( q\right) +sf_\mu ^{\prime }\left( q\right) 
\end{equation}
\noindent where $j_\mu \,$ are the variables used in\cite{minipl} for the
spin and they are such that $j^2=$constant. Introducing (37) in the
Lagrangian of \thinspace \cite{minipl} it is easy to show that: 
\begin{equation}
L_s=\,L(\cite{minipl})\,-\,\,\frac d{dt}\left[ \frac s2\ln \frac{f\cdot \xi }{%
(S\cdot \xi -s^2)}\right] 
\end{equation}
where $\xi _\mu \,$is a constant vector introduced in \cite{minipl}
which satisfies $\xi ^2=-1$. Therefore the two Lagrangians are
dynamically equivalent, though $L_s$ is much simpler. Now two remarks should
be made: First, although we believe that it is possible, we have not been
able to generalize the gauge conditions (29) to the $s\neq 0$ case and
consequently we do not know how to get the intermediate Lagrangian from
the covariant one as we did for the $s=0$ case. Secondly, the canonical
pair $\left( q,\pi \right) \,$is similar to the canonical variables ($%
J^{\left( 0\right) },\varphi $) of \cite{minipl} nevertheless there is no
canonical transformation between these two couples since \thinspace ($%
J^{\left( 0\right) },\varphi $) \thinspace were obtained from the two
independent spin variables by means of a transformation which includes the
momentum $p_\mu \,$differently from our simpler variables.

\section{Summary and Outlook}

By minimally extending the space-time phase space $\left( x_\mu ,p_\nu
\right) \,$ of a spinless massive particle through the introduction of one
couple of canonical variables $\left( q,\pi \right) $ we have derived a
continuous family of Lagrangians $L_s=L_s(\dot x_\mu ,q,\dot q)$ that
describe a free anyon with given mass $m$ and given helicity $\alpha \,\,$.
The Lagrangian $L_s$ has a number of nice features like : correct counting of
degrees of freedom ; canonical spin algebra $(\left\{ S_\alpha ,S_\beta
\right\} =\,\epsilon _{\alpha \beta \gamma }S^\gamma )$ ; canonical space
time structure $(\left\{ x_\alpha ,x_\beta \right\} $ $=0=\left\{ p_\alpha
,p_\beta \right\} ;\,\left\{ x_\alpha ,p_\beta \right\} =\eta _{\alpha \beta
}\,)\,$ and specially remarkable is the simplicity of the case 
where the dual spin vector is light like $\left( S^2=-s^2=0\right) $.
The price we have paid for all those nice features was the loss of explicit
Lorentz covariance, though we have shown that the action obtained is indeed
Lorentz invariant very much like the Floreanini and Jackiw \cite{fj} action
for chiral bosons and the Schwarz and Sen action \cite{ss} for
electrodynamics.

A key ingredient to obtain $L_s$ was the study of realizations of
the $SL(2,\Re )$ algebra in a two dimensional phase space like $\left( q,\pi
\right) \,$. It was shown that , up to a canonical transformation, the
realizations which have the same Casimir can be taken to be of the form
(13) with $f_\mu $ given in (10). Therefore, given $s,\alpha ,$ and $m$
the action $S_s=\int L_sdt$ is physically equivalent ,at the classical
level, for any realization $\tilde S_\mu (\tilde q,\tilde \pi )$ of the $%
SL(2,\Re )$ algebra such that the Casimir is the same, i.e., $\tilde
S^2=S^2=-s^2$.

In section 4 we have shown that the Lagrangian $L_0$ can be alternatively
obtained from the more involved covariant description of anyons of 
\cite{minipl} by means of appropriate gauge conditions followed by a Darboux
transformation. The generalization to the case where $S^2=-s^2\ne 0$ is
still under investigation as well as the search for Lagrangians similar to $%
L_s$ that describe free anyons with time like spin vectors ($S^2>0$).

Finally we mention that the natural sequence for this work is the
quantization of $L_s$. Especially promising is the calculation of the anyon
propagator that although non-covariant may give some clue on the covariant
case and the corresponding field theoretical action for the relativistic
free anyon. Of course, it is also of interest to couple $L_s$ to an external
electromagnetic field whose non-relativistic limit may be important for
practical applications of anyons.

\section{Acknowledgements}

The authors are partially supported by CNPq. This work was also supported by
FAPESP under contracts \#95/4795-8 and \#96/6162-5.


\begin{thebibliography}{99}

\bibitem{Wilczek} F. Wilczek, in Fractional Statistics and Anyon
Superconductivity (World Scientific, Singapore, 1990).

\bibitem{hall} R. B. Laughlin, Phys. Rev. Lett. {\bf 50} (1983) 1395; D.
P. Arovas, in Geometric Phases in Physics, edited by A. Shapere and F.
Wilczek (World Scientific, Singapore, 1989).

\bibitem{super} J. M. Leinaas and J. Myrheim, Nuovo Cimento {\bf B37} (1977)
1; F. Wilczek, Phys. Rev. Lett. {\bf 48} (1982) 1144; {\bf 49} (1982) 957.

\bibitem{schon} J. F. Schonfeld, Nucl. Phys. {\bf B 185} (1981) 157.

\bibitem{bala} A. P. Balachandran {\it et al}, Gauge Symmetries and the
Fibre Bundles, (Springer Verlag - Berlin - 1983).

\bibitem{jn} R. Jackiw and V. P. Nair, Phys. Rev. D {\bf 43} (1991) 1933.

\bibitem{forte} S. Forte, Int. J. Mod. Phys. {\bf A 7} (1992) 1025.

\bibitem{ijpl} M. S. Plyushchay, Int. J. Mod. Phys. {\bf A 7} (1992) 7045.

\bibitem{chou} C. Chou, V. P. Nair and A. P. Polychronakos, Phys. Lett.
{\bf B304} (1993) 105.

\bibitem{gosh} S. Gosh, Phys. Lett. {\bf B 338} (1994) 235.

\bibitem{plb} D. Dalmazi and A. de Souza Dutra, Phys. Lett. {\bf B 343}
(1995) 225.

\bibitem{minipl} J. L. Cortes and M. Plyushchay, Int. J. Mod. Phys. {\bf A
11} (1996) 3331.

\bibitem{ohnu} Y. Ohnuki, in Proceedings of $2^{nd}$ Winter School on
Mathematical Physics, Mt. Sorak, Korea, edited by Y. M. Cho (World
Scientific, Singapore, 1989).

\bibitem{ber} F. A. Berezin and M. S. Marinov, Ann. Phys. (N.Y.) {\bf 104}
(1977) 336.

\bibitem{div} L. Brink, Di Vechia and P. Howe, Nucl. Phys. {\bf B 118}
(1977) 76.

\bibitem{bar} A. Bohm, Y. Ne'eman and A. O. Barut, Dynamical Groups and
Spectrum Generating Algebras, (World Scientific - Singapore - 1988).

\bibitem{fj} R. Floreanini and R. Jackiw, Phys. Rev. Lett. {\bf 59} (1987)
1873.

\bibitem{ss} J. H. Schwarz and S. Sen, Nucl. Phys. {\bf B 411} (1994) 35.


\end{thebibliography}
\end{document}